\documentclass[final,5p,times,twocolumn,preprint]{elsarticle}
\usepackage{amssymb}
\usepackage{amsmath,amssymb,amsfonts}
\usepackage{algorithmic}
\usepackage{graphicx}
\usepackage{textcomp}
\usepackage{xcolor}
\usepackage{wasysym}
\usepackage[strings]{underscore}
\usepackage{xcolor}
\usepackage{pifont}
\usepackage{url}
\usepackage{multirow}
\usepackage{booktabs}
\usepackage{bm}
\usepackage{stfloats}
\usepackage{color}
\usepackage{setspace}
\usepackage{changepage}
\usepackage{threeparttable}
\usepackage{array}
\usepackage{makecell}
\usepackage{apalike}
\journal{XXXXX}

\begin{document}

\begin{frontmatter}

\title{Multi-View Imputation and Cross-Attention Network Based on Incomplete Longitudinal and Multimodal Data for Conversion Prediction of Mild Cognitive Impairment}

\author[1]{Tao Wang}
%\ead{wangtao\_9802@sina.com }

\author[1]{Xiumei Chen}
%\ead{chenxiumei97@163.com}

\author[1]{Xiaoling Zhang}
%\ead{zhangxiaoling9911@163.com}

\author[1]{Shuoling Zhou}
%\ead{zslandsouling@163.com}

\author[1,2,3]{Qianjin Feng\corref{cor1}}
\ead{fengqj99@smu.edu.cn}

\author[1,2,3]{Meiyan Huang\corref{cor1}}
\ead{huangmeiyan16@163.com}

%\cormark[1]

% Corresponding author text
\cortext[cor1]{Corresponding author}

\address[1]{School of Biomedical Engineering, Southern Medical University, Guangzhou 510515, China}
\address[2]{Guangdong Provincial Key Laboratory of Medical Image Processing, Southern Medical University, Guangzhou 510515, China}
\address[3]{Guangdong Province Engineering Laboratory for Medical Imaging and Diagnostic Technology, Southern Medical University, Guangzhou 510515, China}

\begin{abstract}
Predicting whether subjects with mild cognitive impairment (MCI) will convert to Alzheimer's disease is a significant clinical challenge. Longitudinal variations and complementary information inherent in longitudinal and multimodal data are crucial for MCI conversion prediction, but persistent issue of missing data in these data may hinder their effective application. Additionally, conversion prediction should be achieved in the early stages of disease progression in clinical practice, specifically at baseline visit (BL). Therefore, longitudinal data should only be incorporated during training to capture disease progression information. To address these challenges, a multi-view imputation and cross-attention network (MCNet) was proposed to integrate data imputation and MCI conversion prediction in a unified framework. First, a multi-view imputation method combined with adversarial learning was presented to handle various missing data scenarios and reduce imputation errors. Second, two cross-attention blocks were introduced to exploit the potential associations in longitudinal and multimodal data. Finally, a multi-task learning model was established for data imputation, longitudinal classification, and conversion prediction tasks. When the model was appropriately trained, the disease progression information learned from longitudinal data can be leveraged by BL data to improve MCI conversion prediction at BL. MCNet was tested on two independent testing sets and single-modal BL data to verify its effectiveness and flexibility in MCI conversion prediction. Results showed that MCNet outperformed several competitive methods. Moreover, the interpretability of MCNet was demonstrated. Thus, our MCNet may be a valuable tool in longitudinal and multimodal data analysis for MCI conversion prediction. Codes are available at \url{https://github.com/Meiyan88/MCNET}.
\end{abstract}


\end{frontmatter}

\section{Introduction}

Alzheimer’s disease (AD) is characterized by the irreversible impairment of cognitive functions and is one of the most common neurodegenerative diseases in elderly people \citep{alzheimer2022}. Although no effective cure exists for AD, clinical intervention at the early stage may decelerate its progression. Hence, the early diagnosis of AD’s emergence is important for timely treatment to slow down progressive deterioration \citep{zhang2019multi, ning2021relation}. Mild cognitive impairment (MCI) is commonly regarded as a prodromal stage of AD and a critical period for early diagnosis of this disease \citep{scheltens2021alzheimer}. As reported, approximately 9.6\% of subjects who have MCI are expected to progress to AD annually, whereas some MCI subjects maintain a stable clinical condition with the passage of time \citep{mitchell2009rate, ganguli2019mild}. Therefore, the accurate identification of MCI subjects who will progress to AD would be crucial in providing support for delaying AD progression and developing new clinical therapies \citep{zhou2014optimal}. Generally, based on the criteria of whether MCI subjects will convert to AD within three years, subjects can be classified into stable MCI (sMCI) and progressive MCI (pMCI) groups \citep{arco2021data, bucholc2023hybrid}; such classification is denoted as MCI conversion prediction and is the focus in this study.

In clinical practice, the amount of longitudinal and multimodal data is increasing, and has attracted our attention \citep{el2020multimodal}. On the one hand, multimodal neuroimages, such as magnetic resonance imaging (MRI) and positron emission tomography (PET) images, can provide complementary structural and functional information \citep{ferreira2011neuroimaging}. Moreover, extensive studies based on multimodal neuroimages at a single time point have shown reasonable performance in early AD diagnosis and MCI conversion \citep{zhu2021dual}. However, missing data persist in multimodal data, and some useful information may be lost when only modality-complete data are used in MCI conversion or AD predictions \citep{pan2021collaborative, chen2022structure}. Moreover, only a simple concatenation of multimodal features was used in most existing studies, which may bring redundant information and fail to exploit potential associations among different modalities \citep{zhou2019effective}.

On the other hand, AD is a progressive disease, and some early pathological changes, including structural abnormalities within longitudinal variations, can be captured by longitudinal data \citep{zhang2021multi}. Thus, numerous methods have been proposed to use longitudinal data for AD-related analysis and prediction \citep{huang2021imaging, brand2019joint}. Nevertheless, the missing data issue remains a common but great challenge in using longitudinal data. It limits the direct usage of most conventional machine learning or deep learning methods. Besides only using modality-complete and longitudinal data to alleviate this problem, some studies first imputed missing data, and then used the complete data to train a model for AD prediction \citep{che2018recurrent, ghazi2019training}. This decouple two-stage strategy may lead to sub-optimal results, and model performance is heavily influenced by the chosen imputation method \citep{ma2020adversarial}. Many techniques can be used for imputing missing values, such as simple forward/backward filling \citep{lipton2016modeling}, matrix factorization methods based on singular value decomposition \citep{cai2010singular}, statistical methods \citep{el2020multimodal}, and machine learning methods \citep{zhang2021multi}. Recently, recurrent neural networks (RNN), such as long short-term memory (LSTM) \citep{hochreiter1997long} and gated recurrent unit (GRU) networks \citep{cho2014learning}, have been used on data imputation, and advancements have been made \citep{yoon2018estimating}. However, modality-complete data at baseline visit (BL) are required in these methods \citep{ghazi2019training}. Many subjects have no available PET images at BL because of various practical issues (e.g., high cost, poor image quality, and others). Moreover, errors exist in the estimated data, which may affect the performance of subsequent tasks. Therefore, further reducing the errors of imputed data is still a problem to be solved. In clinical practice, MCI conversion prediction should be achieved in the early stages of disease progression to facilitate timely intervention, specifically at BL. Based on this situation, it is desirable for the model to focus on the prediction performance at BL, without requiring longitudinal data as inputs during testing/usage phase. Therefore, how to effectively use disease progression information in longitudinal data, whereas only BL data are required as inputs at the model testing phase is also a problem that need to be considered.

To address the aforementioned challenges, an end-to-end multi-task deep learning framework, named multi-view imputation and cross-attention network (MCNet), was proposed to utilize incomplete longitudinal and multimodal data for MCI conversion prediction (i.e., classify subjects into sMCI and pMCI). The proposed method consists of data imputation and conversion prediction modules. These two modules share the same multimodal features extracted from the RNN-based network for multi-task learning, including data imputation, longitudinal classification, and conversion prediction tasks. First, in the data imputation module, a novel multi-view imputation strategy with adversarial learning was designed to utilize disease progression information to impute MRI/PET data from a longitudinal view and apply the associations between different modalities to impute the PET data from a multimodal view. Moreover, incorporating adversarial learning is conducive to increase the realities and reduce the errors of imputed data. Second, the features obtained from the imputation module were fused by two unique feature fusion blocks, named cross-attention blocks, for final MCI conversion prediction. Various missing data scenarios exist in incomplete longitudinal and multimodal data, which leads to differences in importance among the information contained in the features. Therefore, two cross-attention blocks were developed to weigh features and reflect the information’s importance. Then, the fused features were used to accomplish longitudinal classification and conversion prediction tasks. With the data imputation strategy and cross-attention blocks, the proposed method extracted disease progression information from longitudinal data during training and directly applied the information on BL data to obtain prediction results without feeding longitudinal data in the testing phase. In other words, we aim to use disease progression information that is learned from longitudinal data to improve performance of MCI conversion prediction at a single time point. Based on previous reports, no research has combined disease progression information from longitudinal data and multimodal associations from multimodal data to achieve adversarial multi-view imputation at all time points with small errors and integrated classification and prediction tasks in the same framework to achieve joint optimization for MCI conversion prediction. In summary, the contributions of this work are as follows:

\begin{enumerate}[\textbullet]
	\item Based on incomplete longitudinal and multimodal data, data imputation and MCI conversion prediction were integrated into a unified network and introduced a multi-task learning strategy to achieve joint optimization and improve the prediction performance.
	\item A multi-view imputation strategy was designed for different modalities and time points to achieve data imputation that can cope with various missing data scenarios. Moreover, the adversarial learning was incorporated into the imputation strategy to make an imputation data distribution that was close to the real distribution, thereby further reducing imputation errors.
	\item Two cross-attention blocks were proposed to fuse multimodal features at different time points to capture information importance at different time points and modalities, thereby further improving prediction performance.
	\item With well training, disease progression information learned from longitudinal data can be leveraged by our proposed method to complete MCI conversion prediction with BL data in the testing phase. Hence, our proposed method met the requirement of early prediction in clinical practice. Additionally, our model can still perform well when only single-modal data (e.g., MRI) are available at BL. The proposed method was trained on two datasets provided by the Alzheimer’s Disease Neuroimaging Initiative (ADNI) database (i.e., ADNI-1 and ADNI-2) and tested on two external independent datasets (i.e., ADNI-3 and Open Access Series of Imaging Studies-3 [OASIS-3]). Competitive results were achieved using the proposed method, which further demonstrates the well generalized ability of the proposed method.
\end{enumerate}

\section{Related Work}
\subsection{Multimodal AD-related Analysis}
Many studies have focused on the applications of multimodal neuroimages, which contain complementary information, for AD diagnosis \citep{shi2017multimodal, behrad2022overview}. For example, different deep neural networks were first used to learn high-level features from multimodal data (e.g., PET, MRI, genetic data, and others) at BL. Then, the features learned from the different modalities were concatenated directly for final AD detection \citep{venugopalan2021multimodal}. Zhou~\emph {et al.} \citep{zhou2019effective} proposed a three-stage deep feature learning and fusion framework to accomplish multi-scale feature fusion for AD prediction. Better prediction performance was achieved using multimodal data than single-modal data. However, the direct concatenation of multimodal features may fail to take full advantage of the information from different modalities \citep{ning2021relation}. Zu~\emph{et al.} \citep{zu2018multi} utilized multi-kernel learning to combine multimodal data for AD classification. Furthermore, Leng~\emph {et al.} \citep{LENG2023106788} developed a cross enhanced fusion mechanism to emphasize the correlation and complementarity between multimodal features for AD diagnosis. Although these strategies can be used to effectively combine multimodal data, their effects on the feature fusion of longitudinal and multimodal data need further investigation.

In the longitudinal and multimodal study, associations among different modalities and different time points should be considered. Inspired by the self-attention mechanism \citep{vaswani2017attention}, we introduced two cross-attention blocks to exploit the importance of the features extracted from different modalities at different time points and to enhance the performance of MCI conversion prediction.
\subsection{Longitudinal AD-related Analysis}
An increasing number of studies attempted to utilize these data for AD-related analysis with the increasing amount of available longitudinal data collected at follow-up time points \citep{wang2019ensemble, huang2017longitudinal, zhang2021multi, brand2019joint}. Some studies explored the use of traditional machine learning methods. Huang~\emph {et al.} \citep{huang2021imaging} proposed a novel temporal group sparsity regression and additive model to identify the associations between longitudinal imaging and genetic data for the detection of potential AD biomarkers. More recently, deep learning methods have shown great potential in AD analysis and have been applied to related classification and regression tasks with promising performance \citep{el2020multimodal, jung2021deep}. Among them, RNN-based deep learning methods are often used in longitudinal studies. Nevertheless, conventional RNNs are designed to be used with complete data; incomplete data still present serious problems for the applications of RNN. Some studies tried to alleviate the negative impact of this problem by taking advantage of RNN to deal with variable-length series data for imaging feature extraction and AD diagnosis but directly ignored the missing data issue \citep{huang2021deep}. Che~\emph {et al.} \citep{che2018recurrent} designed a GRU-based method (GRU-D) to introduce a decay mechanism using information on the interval and location of missing values. Then, they combined decay rates with the incomplete longitudinal data to accomplish classification. Moreover, Ghazi~\emph {et al.} \citep{ghazi2019training} proposed a generalized backpropagation through time algorithm for LSTM, and this method can handle missing input and output values. All the missing values of longitudinal data were initialized with zeros.

Although certain attempts have been made in these methods, missing data issue is only considered in longitudinal view, whereas discarded in multimodal view. Hence, further exploration is still needed to reduce the impact of missing data issue on the final prediction task.

\subsection{Data Imputation}
Missing data is a common issue for longitudinal and multimodal data and may decrease the accuracy of MCI conversion prediction. El-Sappagh~\emph {et al.} \citep{el2020multimodal} tried to solve this issue by discarding the subjects with serious missing data conditions and using the \textit{k}-nearest neighbor algorithm to impute missing values for remaining subjects. However, this kind of data imputation method is a decouple two-stage methodology, which may lead to sub-optimal results \citep{ma2020adversarial}. Therefore, some studies trained a unified model to accomplish data imputation and prediction or classification simultaneously. Nguyen~\emph {et al.} \citep{nguyen2020predicting} used the temporal dependencies of RNN to impute a set of longitudinal data and performed classification in a unified model. However, only temporal associations were considered in this method, and the correlations between different modalities at a time point may be ignored. Jung~\emph {et al.} \citep{jung2021deep} integrated data imputation and longitudinal data classification into a unified framework by utilizing information on the interval between missing values, the location of missing values, and multivariate relations, where the relations of different modalities can be reflected by the multivariate relations and reasonable data imputation, and classification results can be achieved by this method. However, estimation errors of imputation data may accumulate in these unified training methods during the feedforward of the RNN \citep{bengio2015scheduled}. Moreover, the missing data issue at BL cannot be addressed in previous studies.

Generative adversarial network has unparalleled advantages in data generation. Therefore, Ma~\emph {et al.} \citep{ma2020adversarial} introduced the adversarial learning strategy into the data imputation and classification framework to solve the accumulated errors problem, which can further improve the classification performance. Inspired by this idea, we also introduced an adversarial loss in the proposed method. This study used such a strategy for the first time in AD longitudinal and multimodal data imputation. Moreover, we designed a novel multi-view imputation method according to the specific scenarios of missing data to effectively impute missing data and solve the missing data issue at BL.

Additionally, these methods were not designed to consider how to complete the prediction using only BL data during the model testing/usage phase. Specifically, most existing methods still require longitudinal data as inputs in the testing/usage phase \citep{el2020multimodal, nguyen2020predicting}. Instead of using longitudinal data during testing phase, the proposed method tried to learn helpful disease progression information from longitudinal data when training and then used the learned information to improve MCI conversion prediction at BL when testing. We hypothesized that even if longitudinal data was limited, the underlying information of disease trajectories can be leveraged by deep learning approaches. When the model was properly trained, the model was able to judge the trend of the disease from BL data through the learned pattern of disease progression and utilized this auxiliary knowledge to enhance the performance of MCI conversion prediction. Therefore, only multimodal data at BL or even single modal data at BL were needed in the testing phase in the proposed method.

\section{Materials}
The brain imaging data used in this paper were obtained from the ADNI (\url{https://www.adni.loni.usc.edu/}) and OASIS-3 databases (\url{https://www.oasis-brains.org/}). A total of 1387 subjects with T1-weighted MRI and fluorodeoxyglucose positron emission tomography (FDG-PET) images in the three ADNI subsets, namely, ADNI-1, ADNI-2, and ADNI-3, were collected in this study. For ADNI-1 and ADNI-2, images at BL and at 6, 12, 24, and 36 months were included when available, whereas only images at BL provided by ADNI-3 were included as the independent testing set. Specifically, subjects that had MRI images at BL and more than two other time points were included, and the available status of PET images was not considered as a criterion for selecting subjects for ADNI-1 and ADNI-2. Moreover, an additional 143 subjects obtained from OASIS-3 were used as another independent testing set. In terms of mini-mental state examination (MMSE) scores and clinical dementia rating, the clinical status of a subject at a time point can be divided into three categories, i.e., cognitive normal (CN), MCI, and AD. In this study, all subjects were further categorized into four groups based on the individual clinical status at BL and future time points, as follows: (a) CN: the subjects were diagnosed as CN at BL and remained CN afterwards; (b) sMCI: the subjects were diagnosed as MCI at all time points; (c) pMCI: the subjects were diagnosed as MCI at BL and then converted to AD within three years; and (d) AD: the subjects had a clinical status of AD at all time points. The number of enrolled subjects and more demographic information are shown in Table \ref{Table 1}. Subjects with reverse conversion of clinical status were removed. The details of the subject number with different image modalities at different time points are listed in Table \ref{Table 2}.

Raw MRI images acquired by 1.5T/3T scanners and PET images preprocessed by ADNI were downloaded. Then, all MRI images were processed through the following procedures \citep{chen2022structure}: (a) anterior commissure-posterior commissure correction by using MIPAV software (\url{https://mipav.cit.nih.gov/}); (b) image intensity inhomogeneity correction by using N4 algorithm; (c) skull stripping via a robust brain extraction network named HD-BET \citep{isensee2019automated}; (d) registering images to Montreal Neurological Institute (MNI) space via advanced normalization tools (ANTs) (\url{https://github.com/ANTsX/ANTs}); (e) segmentation of three main tissues, i.e., gray matter (GM), white matter, and cerebrospinal fluid, by using Atropos algorithm in ANTs; (f) labeling 90 regions-of-interest (ROIs) on all registered images based on the automated anatomical label atlas of MNI space; and (g) computing the GM tissue volume of each ROI in the labeled images. Subsequently, PET images obtained from OASIS-3 were preprocessed in the same way as those obtained from ADNI \citep{jagust2015alzheimer}. Finally, preprocessed PET images were aligned to their corresponding MRI by using co-registration strategy and the average intensity value of each ROI was calculated as a PET feature. Thus, 90-dimensional ROI features were separately extracted from the MRI and PET data for each subject.

\begin{table}[]
	\scriptsize          
	\centering
	\caption{Demographic information of enroalled subjects. The age and MMSE scores are presented by mean ± standard deviation (std).}
	\setlength{\tabcolsep}{2pt}
	\label{Table 1}
	\begin{tabular}{@{}ccccccc@{}}
		\toprule[1.5pt]
		\textbf{Dataset}        & \textbf{Total}       & \textbf{Category} & \textbf{Number} & \textbf Male/Female & \textbf{Age} & \multicolumn{1}{c}{\textbf{\begin{tabular}[c]{@{}c@{}}MMSE\end{tabular}}} \\ \midrule
		\multirow{4}{*}{ADNI-1} & \multirow{4}{*}{543} & CN           & 165             & 90/75           & 75.3 ± 5.2   & 29.0 ± 1.1    \\
		&                      & sMCI         & 144             & 91/53           & 74.6 ± 7.5   & 27.3 ± 1.6    \\
		&                      & pMCI         & 116             & 68/48           & 73.8 ± 6.9   & 26.8 ± 1.8    \\
		&                      & AD           & 118             & 60/58           & 75.1 ± 7.8   & 23.4 ± 1.9    \\
		\multirow{4}{*}{ADNI-2} & \multirow{4}{*}{758} & CN           & 255             & 129/126         & 73.7 ± 5.8   & 29.1 ± 1.2    \\
		&                      & sMCI         & 286             & 163/123         & 71.5 ± 7.5   & 28.2 ± 1.6    \\
		&                      & pMCI         & 104             & 57/47           & 73.4 ± 6.7   & 27.6 ± 1.9    \\
		&                      & AD           & 113             & 69/44           & 74.3 ± 7.8   & 23.8 ± 2.5    \\ \midrule
		\multirow{2}{*}{ADNI-3} & \multirow{2}{*}{86}  & sMCI         & 73              & 44/29           & 75.5 ± 7.6   & 28.5 ± 1.2    \\
		&                      & pMCI         & 13              & 8/5             & 74.0 ± 6.6   & 26.8 ± 2.8    \\ \midrule
		\multirow{4}{*}{OASIS-3}  & \multirow{4}{*}{143} & CN           & 65              & 37/28           & 73.7 ± 8.6   & 29.2 ± 1.0    \\
		&                      & sMCI         & 48              & 27/21           & 74.9 ± 5.6   & 28.4 ± 1.9    \\
		&                      & pMCI         & 23              & 19/4           & 75.6 ± 7.9   & 27.3 ± 2.3    \\
		&                      & AD           & 7              & 4/3             & 76.1 ± 5.6   & 24.4 ± 1.3    \\ \bottomrule[1.5pt]
	\end{tabular}
\end{table}

\begin{table}[]\scriptsize 
	\centering
	%	\captionsetup{labelfont={bf}}
	\caption{Number of available subjects for different modalities at different time points in ADNI-1 and ADNI-2, where M06, M12, M24, and M36 represent 6, 12, 24, and 36 months, respectively.}
	\label{Table 2}
	\setlength{\tabcolsep}{3pt}
	\begin{tabular}{@{}cccccc@{}}
		\toprule[1.5pt]
		\textbf{Dataset} &   \textbf{BL} & \textbf{M06} & \textbf{M12} & \textbf{M24} & \textbf{M36} \\ \midrule
		ADNI-1 (MRI/PET)  & 543/292     & 534/274      & 527/267      & 451/225      & 284/134      \\
		ADNI-2 (MRI/PET)  & 758/606     & 534/0        & 745/111      & 580/297      & 396/1        \\ \bottomrule[1.5pt]
	\end{tabular}
\end{table}

\begin{figure*}[!t]
	\centerline{\includegraphics[scale=0.25]{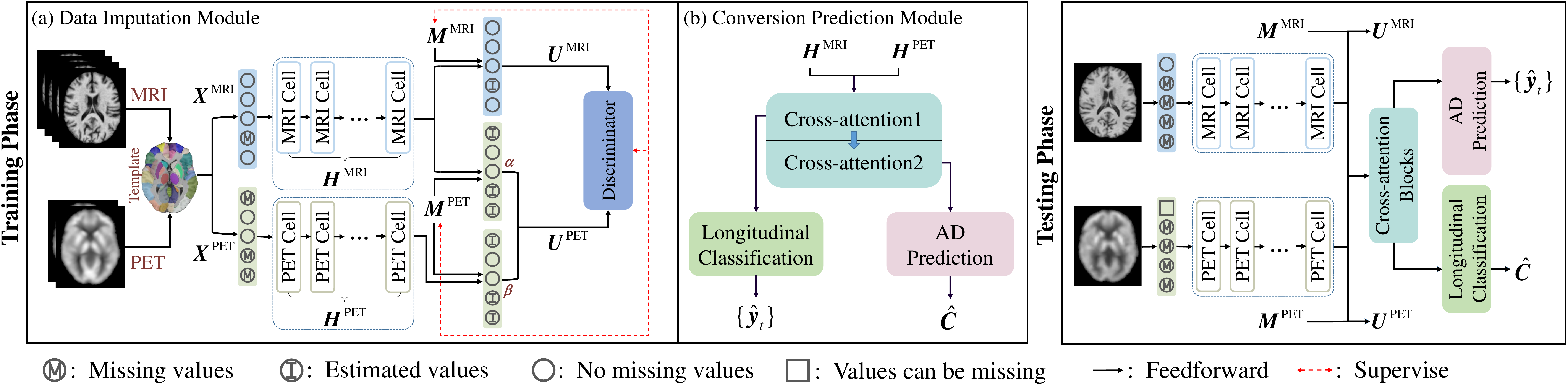}}
	\caption{Overview of the proposed framework. (a) Data imputation module combined with adversarial learning. (b) Conversion prediction module for longitudinal classification and MCI conversion prediction. In the training phase, MRI and PET data at different time points were trained using modules (a) and (b). In the testing phase, only data at BL were used as inputs, where PET data at BL can be available or missing.}
	\label{Figure 1}
\end{figure*}

\begin{table*}[!h]\scriptsize 
	\centering
	%	\captionsetup{labelfont={bf}}
	\caption{Main notations used in this study.}
	\setlength{\tabcolsep}{4pt}
	\label{Table 3}
%	\begin{tabular}{llll}
	\begin{tabular}{llll}
		\toprule[1.5pt]
		\specialrule{0em}{1pt}{1pt}
		\textbf{\textbf{Symbol}}                                                                                        & \textbf{\textbf{Description}}                                                                                                      & \textbf{\textbf{Symbol}}                                                                                                                                                                                                    & \textbf{\textbf{Description}}                                                                                                              \\ \hline\specialrule{0em}{1pt}{1pt} 
		\multicolumn{2}{l}{\textbf{Notations of modality   features}}                                                                                                                                                                                        & $\bm{W}_{{\rm{cs}}}^{{\rm{PET}}}$, $\bm{W}_{{\rm{lg}}}$                                                                                                                                                                     & Learnable weighted coefficients                                                                                                            \\ \cline{1-2} \specialrule{0em}{1pt}{1pt}
		$K$                                                                                                             & Image   modality                                                                                                                   & ${{F}_{{\rm{mini}}}}\left ( \cdot \right )$                                                                                                                                                                                 & The update function of MinimalRNN                                                                                                          \\
		$N$                                                                                                             & Total number of subjects                                                                                                           & ${I} \left ( \cdot \right  )$                                                                                                                                                                                               & The process of data imputation                                                                                                             \\
		$T$                                                                                                             & Total number of time points                                                                                                        & $\left\| \cdot \right\|$                                                                                                                                                                                                    & Mean absolute error                                                                                                                        \\
		$D$                                                                                                             & Dimension of   ROI features                                                                                                        & $\mathcal{L}_{{est}}$                                                                                                                                                                                                       & Estimation loss function                                                                                                                   \\ \cline{3-4} \specialrule{0em}{2pt}{2pt}
		$\bm{x}_{(t)}^{S}$                                                                                              & \begin{tabular}[c]{@{}l@{}}\makebox[4em][s]{ROI features of modality $S$ at $t^{\rm th}$ time point} \\  \end{tabular}                                 & \textbf{Notations in data imputation module}                                                                                                                                                                                &                                                                                                                                            \\
		${\bm{X}^S} = \left \{ \bm{x}_{\rm{(1)}}^S,...,\bm{x}_{(t)}^{S},...,\bm{x}_{(T)}^{S} \right \}$                 & \begin{tabular}[c]{@{}l@{}}\makebox[4em][s]{ROI features of modality $S$ at all time points} \\ \end{tabular}                                          & (c) Notations in adversarial learning                                                                                                                                                                                       &                                                                                                                                            \\ \cline{3-4} 
		${\bm{X} = {\left \{ {\bm{X}^S}\right \}  _{S = \rm{MRI,PET}}}}$                                                & \begin{tabular}[c]{@{}l@{}}\makebox[4em][s]{ROI features of all modalities at all time po-} \\  ints\end{tabular}                                       & $\log \left ( \cdot \right )$                                                                                                                                                                                               & Logarithmic function                                                                                                                       \\
		${\bm{m}_{(t)}^{S}}$                                                                                            & \begin{tabular}[c]{@{}l@{}}\makebox[4em][s]{Mask vector of modality $S$ at $t^{\rm th}$ time point} \\ \end{tabular}                                 & ${{p}_{{\rm{imp}}}}({\bm{u}})$                                                                                                                                                                                              & Imputed data   distribution                                                                                                                \\
		${\bm{M}^S} = \left \{ \bm{m}_{(1)}^S,...,\bm{m}_{(t)}^{S},...,\bm{m}_{(T)}^{S}\right \} $                      & \begin{tabular}[c]{@{}l@{}}\makebox[4em][s]{Mask vectors of modality $S$ at all time poin-} \\ ts\end{tabular}                                          & ${{p}_{{\rm{real}}}}(\bm{u})$                                                                                                                                                                                               & Real data distribution                                                                                                                     \\
		$\bm{y}_{(t)}$                                                                                                  & Longitudinal label at $t^{\rm th}$ timepoint                                                                                       & ${\rm{Ds}}( \cdot )$                                                                                                                                                                                                        & Discriminator                                                                                                                              \\ \specialrule{0em}{1pt}{1pt}
		$\bm{C}$                                                                                                        & Conversion label of a subject                                                                                                      & $\mathcal{L}_{\rm{D}}$                                                                                                                                                                                                      & Discriminator loss function                                                                                                                \\ \cline{1-2} \specialrule{0em}{2pt}{2pt}
		\textbf{Notations in data imputation module}                                                                    &                                                                                                                                    & $\mathcal{L}_{\rm{adv}}$                                                                                                                                                                                                    & Adversarial loss function                                                                                                                  \\  \specialrule{0em}{1pt}{1pt} \cline{3-4} \specialrule{0em}{1pt}{1pt}
		(a) Notations in MinimalRNN                                                                                     &                                                                                                                                    & \textbf{Notations in conversion prediction   module}                                                                                                                                                                        &                                                                                                                                            \\ \cline{1-2} \cline{3-4} \specialrule{0em}{1pt}{1pt}
		$\Phi \left ( \cdot  \right )$                                                                                  & \begin{tabular}[c]{@{}l@{}}A network   for mapping ROI features into a \\ latent representation\end{tabular}                       & $J$                                                                                                                                                                                                                         & \begin{tabular}[c]{@{}l@{}}\makebox[4em][s]{The number of heads in first cross-attention b-} \\  locks\end{tabular}                                             \\
		\textbf{${\bm{z}_{(t)}^S}$}                                                                                     & \begin{tabular}[c]{@{}l@{}}Latent representation of   modality $S$ at $t^{\rm th}$ t- \\ ime point\end{tabular}                      & $J'$                                                                                                                                                                                                                        & \begin{tabular}[c]{@{}l@{}}\makebox[4em][s]{The number of heads in second cross-attentio-} \\  n blocks\end{tabular}                                            \\
		$\bm{g}_{(t)}^S$                                                                                                & Update gate of modality $S$                                                                                                        & $D'$                                                                                                                                                                                                                        & Dimension of hidden features                                                                                                               \\
		${\bm{h}_{{(t)}}^S}$                                                                                            & \begin{tabular}[c]{@{}l@{}}Hidden  feature of modality $S$ at $t^{\rm th}$ time po- \\ int, i.e., output of MinimalRNN\end{tabular} & $\bm{H}_{(t)}$                                                                                                                                                                                                              & \begin{tabular}[c]{@{}l@{}}\makebox[4em][s]{Concatenated hidden features at $t^{\rm th}$ time point} \\ \end{tabular}                                         \\
		$\bm{H}^{S} = \left \{ \bm{h}_{(1)}^{S},...,\bm{h}_{(t)}^{S},...,\bm{h}_{(T)}^{{S}}\right \} $                  & \begin{tabular}[c]{@{}l@{}}\makebox[4em][s]{Hidden   features of modality $S$ at all time po-} \\  ints\end{tabular}                                  & $\bm{Q}_{(t)}^{j}$, $\bm{K}_{(t)}^{j} $, $\bm{V}_{(t)}^{j}$                                                                                                                                                                 & \begin{tabular}[c]{@{}l@{}}Three projection matrices of head $j$ at $t^{\rm th}$ tim- \\ e point in first cross-attention block\end{tabular} \\
		$\bm{W}_{x}^{S}$, $\bm{W}_{h}^{S}$, $\bm{W}_{z}^{S}$, $\bm{b}_{x}^{S}$                                          & \begin{tabular}[c]{@{}l@{}}Learnable weight matrices and bias vectors \\ of modality $S$ in minimalRNN\end{tabular}                & $\bm{A}_{(t)}^{j}$                                                                                                                                                                                                          & \begin{tabular}[c]{@{}l@{}}Attention matrix of head $j$ at $t^{\rm th}$ time point in \\ first cross-attention block\end{tabular}          \\
		$\sigma \left ( \cdot \right )$                                                                                 & Sigmoid activation function                                                                                                        & ${\bm{\tilde H}_{(t)}}$                                                                                                                                                                                                     & Fused features of first   cross-attention block                                                                                            \\
		$\tanh \left ( \cdot \right )$                                                                                  & Hyperbolic tangent function                                                                                                        & $\bm{\tilde H}$                                                                                                                                                                                                             & \begin{tabular}[c]{@{}l@{}}Concatenation of fused features of first cross- \\ attention block\end{tabular}                                  \\
		$\odot$                                                                                                         & Element-wise   product                                                                                                             & $\bm{\hat Q}^{j'}$, $\bm{\hat K}^{j'}$, $\bm{\hat V}^{j'}$                                                                                                                                                                  & \begin{tabular}[c]{@{}l@{}}\makebox[4em][s]{Three projection matrices of head $j'$ in second} \\  cross-attention block\end{tabular}                        \\ \cline{1-2} 
		\textbf{Notations in data imputation module}                                                                    & \textbf{}                                                                                                                          & $\bm{\hat {A}}^{j'}$                                                                                                                                                                                                        & \begin{tabular}[c]{@{}l@{}}\makebox[4em][s]{Attention matrix of head $j'$ in second cross-att-} \\  ention block\end{tabular}                                   \\  
		(b) Notations in multi-view imputation                                                                          & \textbf{}                                                                                                                          & ${\bm{\hat H}}$                                                                                                                                                                                                             & Final features for MCI conversion   prediction                                                                                             \\  \cline{1-2}  \specialrule{0em}{1pt}{1pt}
		$\tilde {\bm{x}}_{{\rm cs},(t)}^{\rm{PET}}$                                                                     & \begin{tabular}[c]{@{}l@{}}\makebox[4em][s]{Estimated ROI features of PET at $t^{\rm th}$ time p-} \\  oint from the multimodal view\end{tabular}       & \begin{tabular}[c]{@{}l@{}}$\bm{W}_{q}^{j}$, $\bm{W}_{k}^{j}$, $\bm{W}_{v}^{j}$, $\bm{W}_{at1}$, $\bm{W}_{at2}$, \\ $\bm{W}_{cls}$, $\bm{W}_{c}$, $\bm{b}_{at1}$, $\bm{b}_{at2}$, $\bm{b}_{cls}$, $\bm{b}_{c}$\end{tabular} & \begin{tabular}[c]{@{}l@{}}Learnable weight matrices and bias vectors in \\ conversion prediction module\end{tabular}                      \\
		$\tilde {\bm{x}}_{{\rm lg},(t)}^{\rm PET}$                                                                      & \begin{tabular}[c]{@{}l@{}}\makebox[4em][s]{Estimated ROI features of PET at $t^{\rm th}$ time p-} \\ oint from the longitudinal view\end{tabular}     & ${\bm{\hat y}_{(t)}}$                                                                                                                                                                                                       & Longitudinal prediction result at $t^{\rm th}$ time point                                                                                  \\
		${\bm{\hat x}}_{(t)}^{S}$                                                                                       & \begin{tabular}[c]{@{}l@{}}Final estimated ROI features of modality $S$ \\ at $t^{\rm th}$ time point\end{tabular}                 & $\bm{\hat {C}}$                                                                                                                                                                                                             & Conversion prediction results                                                                                                              \\
		$\bm{u}_{(t)}^{S}$                                                                                              & \begin{tabular}[c]{@{}l@{}}\makebox[4em][s]{Imputed ROI features of modality $S$ at $t^{\rm th}$ ti-} \\ me point\end{tabular}                         & ${\rm{Softmax}}\left ( \cdot \right )$                                                                                                                                                                                      & Softmax activation function                                                                                                                \\
		$\bm{U}^{S} = \left \{ \bm{u}_{(1)}^{S},...,\bm{u}_{(t)}^{S},...,\bm{u}_{(T)}^{{S}}\right \} $                  & \begin{tabular}[c]{@{}l@{}}\makebox[4em][s]{Imputed ROI features of modality $S$ at all ti-} \\ me points\end{tabular}                                 & $\mathcal{L}_{\rm{cls}}$                                                                                                                                                                                                    & Longitudinal classification loss function                                                                                                  \\
		$\alpha$, $\beta$                                                                                               & Learnable weighted coefficients                                                                                                    & $\mathcal{L}_{\rm{pred}}$                                                                                                                                                                                                   & Conversion prediction loss function                                                                                                        \\
		$\bm{W}_{{\rm{cs}}}^{{\rm{PET}}}$, $\bm{W}_{{\rm{lg}}}$,$\bm{b}_{{\rm{cs}}}^{{\rm{PET}}}$, $\bm{b}_{{\rm{lg}}}$ & \begin{tabular}[c]{@{}l@{}}Learnable weight matrices and bias vectors  \\  in multi-view imputation\end{tabular}                   & $\mathcal{L}$                                                                                                                                                                                                               & Overall loss function                                                                                                                      \\
		${\rm{Concat}} \left ( \cdot \right  )$                                                                         & The operation of  feature concatenation                                                                                            & $\lambda$, $\zeta$, $\xi$                                                                                                                                                                                                   &  \begin{tabular}[c]{@{}l@{}}The hyperparameters in the overall  \\ loss function \end{tabular}                                                                                           \\ \bottomrule[1.5pt]
	\end{tabular}
\end{table*}

\section{Method}

A multi-task learning framework, named MCNet, is proposed for joint data imputation, longitudinal classification, and MCI conversion prediction. Fig. \ref{Figure 1} presents an overview of the proposed method consisting of two modules. In the first module (Fig. \ref{Figure 1} (a)), data imputation was conducted on the MRI and PET ROI features from the multi-views (i.e., longitudinal and multimodal views) using a MinimalRNN-based network. Moreover, an adversarial learning block was proposed to reduce imputation errors and increase the realities of the imputed data. Therefore, longitudinal and multimodal features (hidden features ${\bm{H}}^{\rm{MRI}}$ and ${\bm{H}}^{\rm{PET}}$ in Fig. \ref{Figure 1}) can be well explored with the imputed data. In the second module (Fig. \ref{Figure 1} (b)), two cross-attention blocks were applied to effectively fuse the multimodal and longitudinal features shared from the data imputation module for longitudinal classification and MCI conversion prediction. On the one hand, multimodal features at each time point were fed into the first cross-attention block to fuse multimodal features and capture multimodal associations for longitudinal classification. On the other hand, the fused multimodal features at all time points were fed into the second cross-attention block to exploit potential disease progression information for MCI conversion prediction. The model was trained on all available longitudinal and multimodal data in the training phase, and only data at BL were used as inputs in the testing phase. The proposed method still performed well when PET data was missing at BL.

\subsection{Notations}
In this study, matrices, vectors, and scalars are denoted as bold uppercase letters, boldface lowercase letters, and normal italic letters, respectively. Moreover, all scalars about time points are enclosed in parentheses. The ROI features of MRI and PET data can be represented as ${\bm{X} = {\left \{ {\bm{X}^S}\right \}  _{S = \rm{MRI,PET}}}}$, where $S$ represents the image modality and ${\bm{X}^S} = \left \{ \bm{x}_{\rm{(1)}}^S,...,\bm{x}_{(t)}^{S},...,\bm{x}_{(T)}^{S} \right \}   \in {\mathbb{R}^{{N} \times {T} \times {D}}}$. $N$, $T$, and $D$ are the subject number, number of time points, and dimension of ROI features, respectively. Missing time points often appear in longitudinal MRI and PET data, and mask vectors ${\bm{M}^S} = \left \{ \bm{m}_{(1)}^S,...,\bm{m}_{(t)}^{S},...,\bm{m}_{(T)}^{S}\right \} $ are applied to indicate whether data exist in a time point, where $\bm{m}_{(t)}^{S} \in {\mathbb{R}^{{N} \times {\rm{1}}}}$. In particular, the value in ${\bm{m}_{(t)}^{S}}$ is 1 when $\bm{x}_{(t)}^{S}$ exists and 0 when $\bm{x}_{(t)}^{S}$ is missing. Moreover, the longitudinal and the clinical status labels are denoted as $ \left \{{{\bm y}_{(t)}}\right \}_{{t} = 1}^{T}$ and $\bm{C}$, respectively, where $\bm{y}_{(t)}^{} \in {\mathbb{R}^{{N} \times 1}}$ is longitudinal label at ${t}^{\rm{th}}$ time point. Specifically, if the clinical status is unchanged, ${{\bm{y}_{(t)}} = 0}$; otherwise, ${\bm{y}_{(t)}} = 1$. The longitudinal labels are used in the longitudinal classification task. $\bm{C} \in {\mathbb{R}^{{N} \times {\rm{1}}}}$ indicates the clinical status of a subject. For instance, if the clinical status of a subject is sMCI, then $\bm{C} = 0$. Otherwise, if it is pMCI, then $\bm{C} = 1$. Moreover, $\bm{C}$ is applied to MCI conversion prediction. The main notations used in this study are listed in Table \ref{Table 3}.

\begin{figure}[]
	\centering
	\includegraphics[width=\columnwidth]{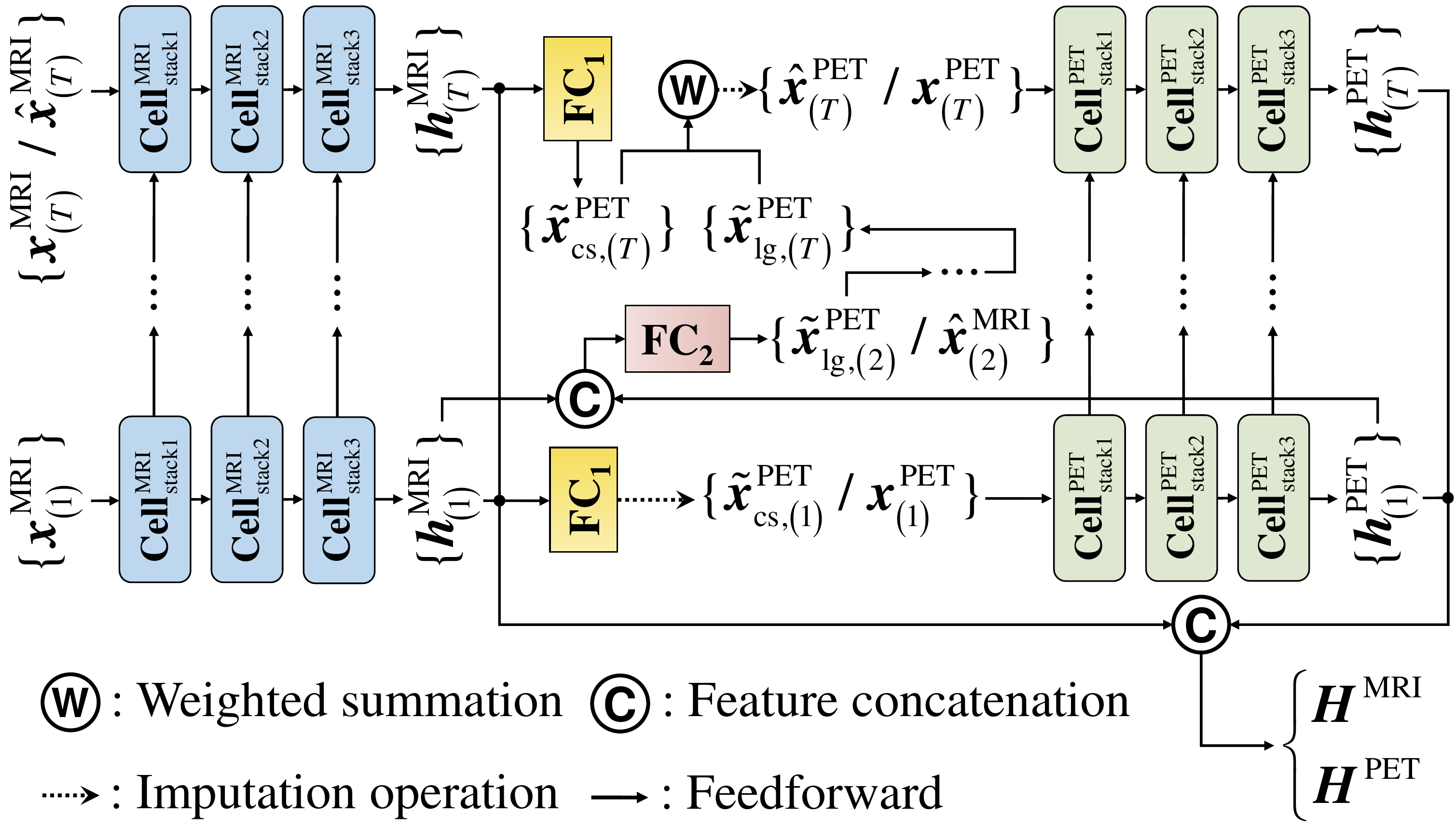}
	\caption{Illustration of data imputation module. The parameters in all time points are shared, i.e., the same memory cells in figure share the same parameters.}
	\label{Figure 2}
\end{figure}

\subsection{Data Imputation Module}
\subsubsection{Minimal Recurrent Neural Network}
We briefly considered MinimalRNN, which is used as the backbone network of the proposed method. MinimalRNN is a distinctive RNN architecture that adopts the minimum number of operations within RNN without sacrificing performance \citep{chen2017minimalrnn}. Moreover, MinimalRNN can be used to capture disease progression information, the trainability of which can also be guaranteed.

A single MinimalRNN is a chain structure composed of several memory cells, where each cell corresponds to a time point. Moreover, all cells share the same parameters. For each cell, the input $\bm{x}_{(t)}^S$ is first fed into a fully connected (FC) network $\Phi \left ( \cdot  \right )$ to generate a latent representation ${\bm{z}_{(t)}^S}$. Through this operation, the representations are confined to move within this latent space \citep{chen2017minimalrnn}. Given $\bm{z}_{(t)}^S$, the weight of update gate ${\bm{g}_{(t)}^S}$ can be simply learned with a sigmoid function $\sigma \left ( \cdot \right )$. Moreover, the update gate $\bm{g}_{(t)}^S$ weighs the contributions of the previous hidden feature ${\bm{h}_{{(t-1)}}^S}$ and the latent representation $\bm{z}_{(t)}^S$ toward the current hidden feature $\bm{h}_{(t)}^S$. Disease progression information can be continuously captured from the data during the forward process from the hidden feature$\bm{h}_{(t-1)}^S$ to $\bm{h}_{(t)}^S$. Therefore, the update process can be formulated as:
\begin{equation}\label{eq1}
{\bm{z}_{(t)}^S} = \tanh\left (\Phi \left ({\bm{x}_{(t)}^S}\right )\right ) = \tanh \left ({\bm{W}_{x}^{S}}{\bm{x}_{(t)}^S} + {\bm{b}_{x}^S}\right )
\end{equation}
\begin{equation}\label{eq2}
{\bm{g}_{(t)}^S} = \sigma \left ({\bm{W}_{h}^{S}}{\bm{h}_{(t-1)}^S} + {\bm{W}_{z}^{S}}{\bm{z}_{(t)}^S}\right )
\end{equation}
\begin{equation}\label{eq3}
{\bm{h}_{(t)}^S} = {\bm{g}_{(t)}^S} \odot {\bm{h}_{(t-1)}^S} + \left (1 - {\bm{g}_{(t)}^S}\right ) \odot {\bm{z}_{(t)}^S}
\end{equation}
where $\bm{W}_{x}^{S}$ and $\bm{b}_{x}^{S}$ denote the parameters of the embedding operation, $\bm{W}_{h}^{S}$ and $\bm{W}_{z}^{S}$ denote the parameters related to update gate $\bm{g}_{(t)}^S$, and $\odot$ is the element-wise product. Finally, the hidden features at all time points can be denoted as $\bm{H}^{S} = \left \{ \bm{h}_{(1)}^{S},...,\bm{h}_{(t)}^{S},...,\bm{h}_{(T)}^{{S}}\right \} $.

\subsubsection{Multi-View Imputation}
In this study, longitudinal and multimodal imaging data were used for accurate MCI conversion prediction. However, the missing data issue from longitudinal and multimodal views is the main limitation when using this type of data. Based on our observations on the ADNI datasets, the cases of missing data can be divided into three main scenarios: (a) PET data missing at BL; (b) MRI data missing at $t^{\rm th}$ time point; and (c) PET data missing at $t^{\rm th}$ time point. We designed a novel multi-view data imputation method that is different from traditional data imputation methods to impute missing values in different cases and address the complexity of the missing data issue.

Compared with longitudinal MRI images, more serious missing data issue appeared in longitudinal PET images. Therefore, as shown in Fig. \ref{Figure 2}, we adopted two separate stacked three-layer MinimalRNNs to capture the different longitudinal features of different modalities and then impute data. MinimalRNN cannot perform feedforward when data are unavailable at BL. Therefore, the missing data issue of PET data at BL (scenario (a)) was considered first. We first imputed the PET data at BL from the multimodal view using a FC network with hyperbolic tangent activation function due to the nonlinear relationship between PET and MRI data. We utilized the hidden feature $\bm{h}_{{(t)}}^{{\rm{MRI}}}$ encoded from the MRI data at BL to estimate the PET data at BL: 
\begin{equation}\label{eq4}
\tilde {\bm{x}}_{{\rm cs},(t)}^{\rm{PET}} = \tanh\left (\bm{W}_{{\rm{cs}}}^{{\rm{PET}}}\bm{h}_{{(t)}}^{{\rm{MRI}}}+ \bm{b}_{{\rm{cs}}}^{{\rm{PET}}}\right )
\end{equation}
where $\bm{W}_{{\rm{cs}}}^{{\rm{PET}}}$, and $\bm{b}_{{\rm{cs}}}^{{\rm{PET}}}$ are learnable parameters of FC network for PET estimation at BL.

For MRI data missing at $t^{\rm th}$ time point (scenario (b)), the potential relationship (i.e., disease progression information) between adjacent time points was utilized for the imputation of the longitudinal view. Previous disease progression information up to $(t-1)^{\rm th}$ time point was contained in the hidden features $\bm{h}_{(t-1)}^{{\rm{MRI}}}$ and $\bm{h}_{(t-1)}^{{\rm{PET}}}$. Accordingly, MRI data can be estimated from the longitudinal view. 
%Meanwhile, the complementary information of different modalities was used for the estimation at $t^{\rm th}$ time point. Meanwhile,  
Specifically, the MRI data $\hat {\bm{x}}_{(t)}^{\rm{MRI}}$ of the $t^{\rm th}$ time point were estimated using the concatenation of hidden features $\bm{h}_{(t-1)}^{{\rm{MRI}}}$ and $\bm{h}_{(t-1)}^{{\rm{PET}}}$. Meanwhile, the PET data $\tilde {\bm{x}}_{{\rm lg},(t)}^{\rm PET}$ of the same time point can also be estimated simultaneously from the longitudinal view. This process can be formulated as:
\begin{equation}\label{eq5}
{\hat {\bm{x}}_{(t)}^{\rm MRI}, \tilde {\bm{x}}_{{\rm lg},(t)}^{\rm PET}} = \bm{W}_{{\rm{lg}}}{\rm{Concat}}\left (\bm{h}_{(t-1)}^{{\rm{MRI}}},\bm{h}_{(t-1)}^{{\rm{PET}}}\right ) + \bm{b}_{{\rm{lg}}} 
\end{equation}
where $\bm{W}_{{\rm{lg}}}$ and $\bm{b}_{{\rm{lg}}}$ are learnable parameters for MRI/PET longitudinal estimation, and ${\rm{Concat}} \left ( \cdot \right  )$ represents feature concatenation. In this study, our focus was solely on imputation from the longitudinal view for MRI data. In clinical practice, an MRI image is typically available at a given time point, wheras the corresponding PET image may be missing from the multimodal view.

For PET data missing at $t^{\rm th}$ time point (scenario (c)), we adaptively combined the two estimated values obtained from the longitudinal and multimodal views to calculate the final estimated value at $t^{\rm th}$ time point. In this way, we leveraged the complementary information from different modalities at the current time point and disease progression information from previous time points for multi-view imputation. Specifically, we can take advantage of the hidden features $\bm{h}_{(t-1)}^{{\rm{MRI}}}$ and $\bm{h}_{(t-1)}^{{\rm{PET}}}$ propagated from the previous $(t-1)$ time points as Eq. (\ref{eq5}) and utilize the hidden feature $\bm{h}_{(t)}^{{\rm{MRI}}}$ encoded from the MRI data at the $t^{\rm th}$ $(t > 1)$ time point as Eq. (\ref{eq4}) to estimate the PET data except for BL:
\begin{equation}\label{eq7}
\left\{
\begin{aligned}
%\nonumber
{\bm{\hat x}}_{(t)}^{{\rm{PET}}} &= {\alpha }\left (\tilde {\bm{x}}_{{\rm cs},(t)}^{\rm{PET}}\right ) + {\beta }\left (\tilde {\bm{x}}_{{\rm lg},(t)}^{\rm PET}\right ), {\quad \rm if \quad} t > 1 \\
{\bm{\hat x}}_{(t)}^{{\rm{PET}}} &= \tilde {\bm{x}}_{{\rm cs},(t)}^{\rm{PET}}, {\quad \rm if \quad} t=1.\\
\end{aligned}
\right.
\end{equation}
where $\alpha$ and $\beta$ are learnable weighted coefficients, and $\alpha + \beta = 1$.

With the estimated PET and MRI data $\hat {\bm{x}}_{(t)}^{S}$ and mask vector ${\bm{m}_{(t)}^{S}}$, we can impute missing data of both modalities at all time points. Specifically, the imputed data $\bm{u}_{(t)}^S$ at $t^{\rm th}$ time point can be defined as:
\begin{equation}\label{eq6}
\bm{u}_{(t)}^{S} = \bm{m}_{(t)}^{S} \odot \bm{x}_{(t)}^{S} + \left (1 - \bm{m}_{(t)}^{S}\right )  \odot \hat {\bm{x}}_{(t)}^{S}
\end{equation}

After the imputation steps, the update equation of MinimalRNN can be formulated as:
\begin{equation}\label{eq8}
\begin{split}
\bm{h}_{(t)}^{S} &= {{F}_{{\rm{mini}}}}\left (\bm{h}_{(t-1)}^{S},{I}\left (\bm{x}_{(t)}^{S},\bm{m}_{(t)}^{S},\bm{h}_{(t-1)}^{S}\right ) \right )  \\
&= {{F}_{{\rm{mini}}}}\left (\bm{h}_{(t-1)}^{S}, \bm{u}_{(t)}^{S}\right ) 
\end{split}
\end{equation}
where ${I} \left ( \cdot \right  )$ represents the process of data imputation, and ${{F}_{{\rm{mini}}}}\left ( \cdot \right )$ represents the update function of MinimalRNN as listed in Eqs. (\ref{eq1})-(\ref{eq3}).

Therefore, we can obtain the complete longitudinal MRI and PET data $\bm{U}^{S} = \left \{ \bm{u}_{(1)}^{S},...,\bm{u}_{(t)}^{S},...,\bm{u}_{(T)}^{{S}}\right \} $ and corresponding hidden features $\bm{H}^{S}$ through the imputation module. Finally, mean absolute error (MAE) was used to measure the loss between the estimated data and the real data, which can be defined as:
\begin{equation}\label{eq9}
\mathcal{L}_{{est}}^{} = \sum\limits_{t = 1}^T \sum\limits_S^{\left \{ {\rm{MRI}},{\rm{PET}}\right \}  } \left ( {{{\left\| {\bm{x}_{(t)}^S -  {\bm{\hat x}}_{(t)}^{S}} \right\|}_1} \odot \bm{m}_{(t)}^S} \right )
\end{equation}

\subsubsection{Adversarial Learning}
Although multi-view estimation was applied to the estimation of longitudinal and multimodal ROI features, some estimation errors still existed. These errors may be accumulated in the feedforward of MinimalRNN. Thus, an adversarial learning strategy was incorporated into the proposed method to alleviate this dilemma. The adversarial learning block can be defined as a minimax game. Our goal was to learn an imputed data distribution ${{p}_{{\rm{imp}}}}({\bm{u}})$ that matched the real data distribution ${{p}_{{\rm{real}}}}(\bm{u})$. 

Specifically, we added a discriminator consisting of a multilayer perceptron (MLP) with a sigmoid function. The primary objective of this addition was to enforce the close approximation of ${{p}_{{\rm{imp}}}}({\bm{u}})$ to ${{p}_{{\rm{real}}}}(\bm{u})$ by fooling the discriminators, thereby mitigating the negative impact of missing values. The supervision signal was provided by mask vectors. Thus, the discriminator loss can be defined as follows:
%\begin{small}
\begin{equation}\label{eq10}
\begin{split}
\mathcal{L}_{\rm{D}} = -& \left [\mathbb{E}_{\bm{x}\sim{{p}_{{\rm{real}}}}(\bm{u})}\log \left ({\rm{Ds}}\left (\bm{x} \right ) \right )+\mathbb{E}_{\hat {\bm{x}}\sim{{p}_{{\rm{imp}}}} \left ( {\bm{u}} \right )}\log \left (1-{\rm{Ds}}(\hat {\bm{x}}) \right ) \right] \\
= - & \sum_{t=1}^{T}  \sum_{S}^{\left \{ \rm MRI,PET \right \}  } \bm m^S_{(t)} \odot \log \left ( {\rm Ds}\left ({\bm u}_{(t)}^{S}  \right )  \right )  \\ - & \sum_{t=1}^{T}  \sum_{S}^{\left \{ \rm MRI,PET \right \}  } \left ( 1- \bm m^S_{(t)} \right )  \odot \log \left ( 1-{\rm Ds}\left ( {\bm u}_{(t)}^{S}  \right )  \right )    
\end{split}
\end{equation}
where ${\rm{Ds}}( \cdot )$ denotes the discriminator function and its output is the estimated mask probability. Therefore, the estimated probability for the real data should be maximized to 1, and the estimated probability for the imputed data should be minimized to 0. Then, we introduced an adversarial loss in the data imputation stage to let MinimalRNN maximize the probability of the discriminator output, which will be backpropagated to further optimize the parameters of the MinimalRNN:

\begin{equation}\label{eq11}
%\fontsize{20pt}{skip}
\mathcal{L}_{{\rm{adv}}} = \sum_{t=1}^{T}  \sum_{S}^{\left \{ \rm MRI,PET \right \}  } \left ( 1- \bm m^S_{(t)} \right )  \odot \log \left ( 1-{\rm Ds}\left ( {\bm u}_{(t)}^{S}  \right )  \right ) 
\end{equation}
%\end{small}

Thus, our model first updated the discriminators ${\rm{Ds}}( \cdot )$ to distinguish the real data from the imputed data with $\mathcal{L}_{\rm{D}}$ and then updated MinimalRNNs with $\mathcal{L}_{\rm{adv}}$. Notably, we considered that the case of missing data in the testing phase as an extreme scenario with only BL data. By leveraging the disease progression information and multimodal correlations learned during training phase, as well as the proposed imputation strategy, we can obtain longitudinal and multimodal features for prediction based solely on the available BL data.

\subsection{Conversion Prediction Module}

Our main goal was to perform MCI conversion prediction, that is, to classify subjects into sMCI and pMCI at BL using the proposed method. Thus, we designed a conversion prediction module to capture the longitudinal and multimodal associations and then developed two cross-attention blocks to fuse the longitudinal and multimodal features effectively. Data imputation module and conversion prediction modules share the same features extracted by MinimalRNNs. In this way, we can simply implement a multi-task learning strategy. We accomplished one of the tasks (i.e., data imputation task) in the data imputation module. In this module, besides MCI conversion prediction, we added another auxiliary task (i.e., longitudinal classification) to determine whether the clinical status of the subjects had changed at each time point. Specifically, this task can be used to exploit predictive representation $\bm{h}_{(t)}^S$ and help the training of the first cross-attention block at each time point, which can contribute to improving the performance of MCI conversion prediction. In conversion prediction module, we developed two cross-attention blocks for feature fusion to effectively combine longitudinal and multimodal information. Specifically, the first cross-attention block was mainly used to explore the relationships among different modalities and fuse the multimodal features of each time point, to determine the importance of different modalities for longitudinal classification/conversion prediction tasks; the second cross-attention block was mainly used to investigate the importance of fused multimodal features from different time points for MCI conversion prediction, and fuse the multimodal features of all time points.

In the first cross-attention block, the hidden features were fused from the multimodal view at each time point through the self-attention mechanism, which was based on a multi-head attention strategy. For head $j$, the concatenated hidden features $\bm{H}_{(t)} = {\rm Concat} \left ( \bm{h}_{(t)}^{{\rm{MRI}}},\bm{h}_{(t)}^{{\rm{PET}}} \right ) \in {\mathbb{R}^{{N} \times 2 \times {D'}}}$ at $t^{\rm th}$ time point was first translated to three matrices, namely, $\bm{Q}_{(t)}^{j} \in {\mathbb{R}^{{N} \times 2 \times ({D'}/{J})}}$, $\bm{K}_{(t)}^{j} \in {\mathbb{R}^{{N} \times 2 \times ({D'}/{J})}}$, and $\bm{V}_{(t)}^{j} \in {\mathbb{R}^{{N} \times 2 \times ({D'}/{J})}}$, with three projection matrices (i.e., $\bm{W}_{q}^{j},\bm{W}_{k}^{j},\bm{W}_{v}^{j} \in {\mathbb{R}^{{D'} \times ({D'}/{J})}}$ for all subject), where $J$ is the number of heads, and $D'$ is dimension of hidden features. Then, the attention matrices ${\bm{A}_{(t)}^j}$ of different heads can be calculated with $\bm{Q}_{(t)}^j$, $\bm{K}_{(t)}^j$, and $\bm{V}_{(t)}^j$ at each head as follows:
\begin{equation}\label{eq12}
\bm{A}_{(t)}^{j} = {\rm{softmax}}\left (\bm{Q}_{(t)}^{j}{{\bm{K}_{(t)}^{j}}^{\rm T}} \big / \left (\sqrt {{D'}\big / {J} } \right ) \right )\bm{V}_{(t)}^{j}
\end{equation}
Next, all heads were concatenated together on feature dimensions and fed into a FC layer to obtain the residual features, which were used to add to the original features $\bm{H}_{(t)}$ to obtain fused features:
\begin{equation}\label{eq14}
{\bm{\tilde H}_{(t)}} = \left ( {\bm{W}_{\rm{at_{1}}}}{\rm{Concat}}\left (\bm{A}_{(1)}^1...,\bm{A}_{(t)}^{j},...,\bm{A}_{(T)}^{J} \right ) + {\bm{b}_{\rm{at_{1}}}} \right ) + \bm{H}_{(t)}
\end{equation}
where $\bm{W}_{\rm{at_{1}}} \in {\mathbb{R}^{ D' \times D'}}$ and $\bm{b}_{\rm{at_{1}}} \in {\mathbb{R}^{{D'} \times 1}}$ are the parameters of the FC layer for the concatenated attention matrices. Moreover, all subjects and both modalities shared the same $\bm{W}_{\rm{at_{1}}}$ and $\bm{b}_{\rm{at_{1}}}$. The fused features can be used to classify whether the clinical status changed according to ${\bm{\tilde H}_{(t)}}$ at $t ^{\rm th}$ time point:
\begin{equation}\label{eq15}
{\bm{\hat  y}_{(t)}} = {\rm{softmax}}\left (\bm{W}_{{\rm{cls}}}{\bm{\tilde H}_{(t)}} + \bm{b}_{{\rm{cls}}}\right )
\end{equation}
%\begin{spacing}{1.0}
where $\bm{W}_{{\rm{cls}}}$ and $\bm{b}_{{\rm{cls}}}$ are the learnable parameters for MRI and PET longitudinal classification, and ${\rm{Softmax}}\left ( \cdot \right )$ denotes softmax activation function. The universal cross-entropy loss for this task is defined as:
%\end{spacing}
%\begin{small}
\begin{equation}\label{eq16}
\begin{split}
\mathcal{L}_{{\rm{cls}}} = - \sum\limits_{t=1}^{T}{\bm{m}_{(t)}^{{\rm{MRI}}}} \odot \left (\bm{y}_{(t)}\log \left (\hat{\bm{y}}_{(t)}\right )  +  \left (1-\bm{y}_{(t)} \right )\log\left (1-\hat{\bm{y}}_{(t)}\right ) \right )
\end{split}
\end{equation}
%\end{small}
where ${\bm{y}}_{(t)}$ and $\hat {\bm{y}}_{(t)}$ are the true and estimated probabilities of a clinical status’ change at $t ^{\rm th}$ time point.

Then, conversion prediction task was incorporated into the proposed method. The head number of the second cross-attention block was set as $J'$. Specifically, the features in all time points were concatenated to prevent the loss of disease progression information at the early time points, where the concatenated features $\bm{\tilde H} = {\rm Concat} \left ( \tilde {\bm{H}}_{(1)},...,\tilde {\bm{H}}_{(t)},...,\tilde {\bm{H}}_{(T)}\right ) \in {\mathbb{R}^{{N} \times T \times 2{D'}}}$ were fed into the second cross-attention block to integrate longitudinal and multimodal information. Similar to the first cross-attention block, three matrices, namely, $\bm{\hat Q}^{j'} \in {\mathbb{R}^{{N} \times T \times ({2D'}/{J'})}}$, $\bm{\hat K}^{j'}\in {\mathbb{R}^{{N} \times T \times ({2D'}/{J'})}}$, and $\bm{\hat V}^{j'} \in {\mathbb{R}^{{N} \times T \times ({2D'}/{J'})}}$, were used to calculate the attention matrix $\bm{{\hat A}}^{j'}\in{\mathbb{R}^{{N} \times T \times ({2D'}/J')}}$ in head $j'$. After concatenating the attention matrices from $J'$ heads, the concatenated attention matrix was fed into a FC layer with parameters $\bm{W}_{\rm{at_{2}}}\in{\mathbb{R}^{ 2D' \times 2D'}}$ and $\bm{b}_{\rm{at_{2}}}\in{\mathbb{R}^{2D' \times 1}}$, and then final features ${\bm{\hat H}} \in {\mathbb{R}^{{N} \times T \times 2{D'}}}$ for MCI conversion prediction were obtained. Hence, the prediction results were defined as:
\begin{equation}\label{eq17}
{\bm{\hat {C}}} = {\rm{softmax}}\left (\bm{W}_{{\rm{c}}} {\hat{\bm H}} + \bm{b}_{{\rm{c}}}\right )
\end{equation}
where ${\bm{W}_{{\rm{c}}}}$ and ${\bm{b}_{{\rm{c}}}}$ are learnable parameters for MCI conversion prediction. The class imbalance in subjects was serious. Thus, focal cross-entropy loss was applied:
\begin{equation}\label{eq18}
\mathcal{L}_{{\rm{pred}}} = - {\mu }{\left (1 - \hat {\bm{C}}\right )^{\gamma }}\log \left (\hat {\bm{C}}\right )
\end{equation}
where $\mu$ and $\gamma$ are set to 0.3 and 2, respectively.

The overall loss function of our proposed method can be defined as follows:
\begin{equation}\label{eq19}
\mathcal{L} = {\lambda }\mathcal{L}_{\rm{est}} + {\zeta }\mathcal{L}_{\rm{adv}} + {\xi }({\rm{ }}\mathcal{L}_{\rm{cls}} + \mathcal{L}_{{\rm{pred}}})
\end{equation}
where $\lambda$, $\zeta$, and $\xi$ are hyperparameters. Hence, our model can be trained in an end-to-end manner, and joint optimization for data imputation, longitudinal classification, and MCI conversion prediction can be achieved.

\section{Experiment}
\subsection{Experimental Settings}
In this study, longitudinal and multimodal ROI features were used to evaluate the prediction and imputation performances of the proposed method. As described in the \textit{Materials} section, three ADNI subsets, as well as the OASIS-3 database, were enrolled in our experiments. Moreover, the proposed method was implemented using PyTorch, and all experiments were performed on a server with NVIDIA TITAN X (Pascal) GPU.

In the experiments, a hold-out method was used, and all subjects from the ADNI-1 and ADNI-2 datasets were partitioned into 10 non-overlapping subsets with the same proportion of each class. Among which, eight subsets were applied for training, one was utilized for validation, and one was used for testing. For subjects in the training set, data at all available time points were used to train the networks, whereas only data at BL were used to select the hyperparameters and evaluate the networks for subjects in the validation and testing sets. The data partitioning process was repeated five times, and the results of the validation and testing sets were achieved in each process. The final results for the ADNI-1 and ADNI-2 datasets were obtained from the average of five results in the testing set. The subjects in ADNI-3 had FDG-PET scans at BL but had PET scans with other tracers (e.g., Pittsburgh compound B) at subsequent time points, which means that only the PET data at BL were available in ADNI-3 for testing. Different from ADNI-3, OASIS-3 contained a certain amount of longitudinal FDG-PET data. However, in all subjects containing longitudinal MRI and PET data, subjects that belong to the MCI category were lacking. According to the characteristics of different datasets, they were used in the different experiments for performance assessment. See Table \ref{Table 4} for details.

\begin{table}[!t]\scriptsize 
	\centering
	\caption{Details of different testing sets. }
	\setlength{\tabcolsep}{2pt}
	\label{Table 4}
	\begin{adjustwidth}{-0.15cm}{}
		\resizebox{9cm}{!}{
			\begin{tabular}{@{}cccc@{}}
				\toprule[1.5pt]
				\textbf{Denotation} & \textbf{Included Data}                & \textbf{Usage}              & \textbf{Subject Number} \\ \midrule
				ADNI-1/2-A          & Longitudinal data (incomplete BL data) & Section 5.3.1         & 130 MCI                  \\
				ADNI-1/2-C          & Longitudinal data (complete BL data)   & Section 5.3.2  & 81 MCI                   \\
				ADNI-3-A            & BL data (only MRI data)               & Section 5.3.1       & 86 MCI                   \\
				ADNI-3-C            & BL data (complete MRI and PET data)   & Section 5.3.2  & 86 MCI                   \\
				OASIS-3-A           & BL data (only MRI data)               & Section 5.3.1        & 65 MCI                   \\
				OASIS-3-C           & Longitudinal data (complete BL data)   & Section 5.3.2 & 65 CN / 6 MCI / 7 AD         \\ \bottomrule[1.5pt]
			\end{tabular}
		}
	\end{adjustwidth}
\end{table}

\subsection{Implementation Details}
The proposed method with five different losses was optimized from three different stages to balance the computational costs and prediction accuracy. Adam optimizer was used during the training process, and a $\ell 2$-regularization was applied to avoid overfitting. The optimized hyperparameters were selected according to the best average value of area under receiver operating characteristic curve (AUC) on the validation set at each step. (a) Data imputation module was first trained separately according to the data imputation task and corresponding losses listed in Eqs. (\ref{eq9})-(\ref{eq11}). The hyperparameters associated with network structure were determined, and the layer number and dimension of hidden features in MinimalRNN were set to 3 and 128, respectively. Moreover, some hyperparameters in Eq. (\ref{eq19}) were also determined, where $\lambda$ and $\zeta$ were set to 2 and 10, respectively. (b) Conversion prediction module was then trained according to the longitudinal and conversion prediction tasks. The parameters of the two cross-attention blocks were optimized in this step, and the numbers of heads in two cross-attention blocks was set 4. Besides, the data used in the first two steps include all available time points. (c) The two modules were integrated, and end-to-end optimization was achieved using data at BL. In this step, all hyperparameters were finalized: $\xi$ was 10 in Eq. (\ref{eq19}), learning rate was set to $5 \times 10^{-3}$, and weight decay was set to $5 \times 10^{-4}$. The search space of different hyperparameters and the optimized hyperparameters selected in our method are summarized in Table \ref{Table 6}. When all hyperparameters were determined, the model was retrained by an end-to-end manner.

Several quantitative metrics were used to evaluate the methods’ performance in different tasks. Accuracy (ACC), AUC, and balanced accuracy (BAC) were applied for prediction task, and MAE and root mean square error (RMSE) were used for the quantitative evaluation of the imputation task. AUC represents the probability that the predicted positive samples are ranked before the negative samples \citep{huang2005using}. It is suitable for handling class imbalance problem in the datasets. Thus, paired \textit{t}-test (at 95\% significance level) was conducted for statistical significance test in the prediction task. 

\begin{table}[!t]\scriptsize 
	\centering
	%	\captionsetup{labelfont={bf}}
	\caption{Hyperparameter search space and selected hyperparameters.}
	\setlength{\tabcolsep}{2pt}
	\label{Table 6}
	\begin{adjustwidth}{-0.15cm}{}
		\resizebox{9cm}{!}{
			\begin{tabular}{@{}ccc@{}}
				\toprule[1.5pt]
				\textbf{Hyperparameters}    & \textbf{Search space}             & \textbf{Selected value} \\ \midrule
				MinimalRNN layers number    & {[}$1$, $2$, $3$, $4${]}                  & $3$                 \\
				MinimalRNN hidden dimension & {[}$64$, $128$, $256$, $512${]}           & $128$               \\
				Heads number $J$              & {[}$2$, $4$, $6$, $8${]}                  & $4$                 \\
				Heads number $J'$              & {[}$2$, $4$, $6$, $8${]}                  & $4$                \\
				${\lambda }$                & {[}$0.1$, $1$, $2$, $10$, ${10^{2}, 10^{3}}${]}     & $2$                 \\
				${\zeta }$                  & {[}$0.1$, $1$, $2$, $10$, ${10^{2}, 10^{3}}${]}     & $10$                \\
				${\xi }$                    & {[}$1$, $2$, $10$, ${10^{2}, 10^{3}, 10^{4}}${]}     & ${10}$               \\
				Learning rate               & {[}${5 \times 10^{-5} }, {5 \times 10^{-4} }, {5 \times 10^{-3} }, {5 \times 10^{-2} }, {5 \times 10^{-1} }${]} & ${5 \times 10^{-3} }$           \\
				Weight decay                & {[}${5 \times 10^{-6} }, {5 \times 10^{-5} }, {5 \times 10^{-4} }, {5 \times 10^{-3} }${]} & ${5 \times 10^{-4} }$            \\ 
				\bottomrule[1.5pt]
			\end{tabular}
		}
	\end{adjustwidth}
\end{table}

\begin{table}[!t] 
	\centering
	%	\captionsetup{labelfont={bf}}
	\caption{Ablation experiments of the proposed method on different datasets for MCI conversion prediction. Abbreviations: CB, cross-attention blocks; AL, adversarial learning; DI, data imputation; LC, longitudinal classification. The results of ADNI-1/2-A are reported as mean ± std.}
	\label{Table 5}
	\setlength{\tabcolsep}{3pt}
	%\begin{adjustwidth}{-0.5cm}{}
	
	\begin{threeparttable}
		\resizebox{9cm}{!}{
			\begin{tabular}{@{}ccccccccccccc@{}}
				\toprule[2pt]
				\multirow{2}{*}{\textbf{LC}} & \multirow{2}{*}{\textbf{CB}} & \multirow{2}{*}{\textbf{DI}} & \multirow{2}{*}{\textbf{AL}} & \multicolumn{3}{c}{\textbf{ADNI-1/2-A}}                                     & \multicolumn{3}{c}{\textbf{ADNI-3-A}}                & \multicolumn{3}{c}{\textbf{OASIS-3-A}}                \\
				&                     &                     &                     & ACC                    & AUC                    & BAC                    & ACC            & AUC            & BAC            & ACC            & AUC            & BAC            \\ \midrule
				$\times$                   & $\surd$                   & $\surd$                   & $\surd$                   & 0.813 ± 0.023          & 0.832 ± 0.038          & 0.795 ± 0.024          & 0.802          & 0.806          & 0.789          & 0.808          & 0.825          & 0.772          \\
				$\surd$                   & $\times$                   & $\surd$                   & $\surd$                   & 0.815 ± 0.014          & 0.826 ± 0.030          & 0.808 ± 0.011          & \textbf{0.813} & 0.819          & 0.802          & 0.817          & 0.820          & 0.798          \\
				$\surd$                   & $\surd$                   & $\times$                   & $\surd$                   & 0.802 ± 0.029          & 0.818 ± 0.026          & 0.798 ± 0.027          & 0.756          & 0.765          & 0.761          & 0.817          & 0.806          & 0.786          \\
				$\surd$                   & $\surd$                   & $\surd$                   & $\times$                   & 0.812 ± 0.030          & 0.824 ± 0.024          & 0.803 ± 0.026          & 0.791          & 0.821          & 0.782          & 0.798          & 0.838          & \textbf{0.801} \\
				$\surd$                   & $\surd$                   & $\surd$                   & $\surd$                   & \textbf{0.830 ± 0.019} & \textbf{0.842 ± 0.032} & \textbf{0.813 ± 0.032} & 0.802          & \textbf{0.849} & \textbf{0.820} & \textbf{0.827} & \textbf{0.857} & 0.799          \\ \bottomrule[2pt]
			\end{tabular} 
		}
	\end{threeparttable}            
\end{table}

\subsection{Experimental Results and Analysis}
\subsubsection{Ablation Study}
In this section, each of the components in the proposed method was removed separately to investigate its influence on the prediction performance of the proposed method. The results of ablation experiments are shown in Table \ref{Table 5}. All ablation experiments were conducted on the ADNI-1 and ADNI-2 testing sets (ADNI-1/2-A), and two other independent testing sets (ADNI-3-A and OASIS-3-A).

First, discarding longitudinal classification brought performance degradation in MCI conversion prediction, which indicates that the multi-task learning strategy is useful for MCI conversion prediction. Second, the removal of cross-attention blocks led to the direct concatenation of multimodal features at different time points. The worse result implies that considering the relationships among different modalities at different time points plays an important role in MCI conversion prediction. Third, when the data imputation task (i.e., the whole data imputation module) was discarded, all missing data were imputed with mean values, and the corresponding imputation loss listed in Eq. (\ref{eq9}) was removed. The performance of MCI conversion prediction decreased obviously, which proves the effectiveness of using a unified framework for imputation and prediction. Fourth, the performance of MCI conversion prediction declined when the adversarial learning module was ignored. Moreover, the imputation errors were tested on ADNI-1/2-A after the removal of adversarial learning and compared with the imputation errors produced by the proposed method. The proposed method achieved a 0.049 and 0.063 decrease in MAE and RMSE, respectively, which proves that the adversarial learning module can help further reduce imputation errors and improve prediction performance.

In summary, the proposed method achieved the best prediction performance, indicating that the proposed components are useful for MCI conversion prediction. Moreover, the proposed method achieved AUCs of 0.849 and 0.857 on two independent testing sets under the situation of using only MRI, which demonstrates that imputation from the multimodal view can effectively ensure the prediction performance when only MRI at BL is used. Moreover, the result also proves that our method is flexible in data requirements and can achieve reasonable performance without PET data.

\begin{table*}[htbp]\scriptsize 
	\centering
	%	\captionsetup{labelfont={bf}}
	\caption{Imputation errors and prediction performance of different methods. The results of ADNI-1/2-C are reported as mean ± std, and * denotes significant difference with \textit{p}-value $\textless$ 0.05. Data imputation cannot be implemented in GRU-D, SVM-based method and MLP-based method, so metrics related to imputation performance in these methods are represented by `--'.}
	\setlength{\tabcolsep}{3pt}
	\label{Table 7}
	%\begin{adjustwidth}{-0.8cm}{}
	
	\begin{threeparttable}
		\resizebox{18.5cm}{!}{
			\begin{tabular}{@{}ccccccccccccccc@{}}
				\toprule[1.5pt]
				\multirow{2}{*}{\textbf{Method}} & \multicolumn{7}{c}{\textbf{ADNI-1/2-C}}                                                                                                                                      & \multicolumn{4}{c}{\textbf{OASIS-3-C}}                            & \multicolumn{3}{c}{\textbf{ADNI-3-C}}            \\
				& MAE(MRI)               & RMSE(MRI)              & MAE(PET)               & RMSE(PET)              & ACC                    & AUC                    & BCA                    & MAE(MRI)       & RMSE(MRI)      & MAE(PET)       & RMSE(PET)      & ACC            & AUC            & BCA            \\ \midrule
				SVM-based                      & --                      & --                      & --                      & --          & 0.732 ± 0.023          & 0.777 ± 0.024          & 0.723 ± 0.043          & --              & --              & --              & --          & 0.720          & 0.753          & 0.697          \\
				MLP-based                      & --                      & --                      & --                      & --          & 0.766 ± 0.027          & 0.790 ± 0.024          & 0.738 ± 0.033          & --              & --              & --              & --         & 0.773          & 0.761          & 0.755          \\
				LSTM-Robust                      & 0.686 ± 0.050          & 1.010 ± 0.065          & 0.815 ± 0.101          & 1.049 ± 0.130          & 0.807 ± 0.021          & 0.774 ± 0.021          & 0.774 ± 0.021          & 0.807          & 1.027          & 1.016          & 1.321          & 0.760          & 0.768          & 0.768          \\
				GRU-D                            & --                      & --                      & --                      & --                      & 0.803 ± 0.038          & 0.816 ± 0.031          & 0.795 ± 0.026          & --              & --              & --              & --              & 0.773          & 0.819          & 0.775          \\
				AJRNN                            & 0.363 ± 0.043          & 0.511 ± 0.067          & 0.428 ± 0.036          & 0.551 ± 0.046          & 0.831 ± 0.024          & 0.830 ± 0.016          & 0.810 ± 0.017          & 0.398          & \textbf{0.517}          & 0.652          & 0.746          & 0.800          & 0.829          & \textbf{0.828} \\
				DRM                              & 0.403 ± 0.042          & 0.574 ± 0.042          & 0.530 ± 0.015          & 0.699 ± 0.015          & 0.821 ± 0.025          & 0.819 ± 0.030          & 0.819 ± 0.030          & 0.588          & 0.710          & 0.781          & 0.891          & \textbf{0.813} & 0.818          & 0.813          \\
				MCNet-Linear                      & 0.379 ± 0.078          & 0.598 ± 0.077          & 0.471 ± 0.055          & 0.607 ± 0.048          & 0.828 ± 0.012          & 0.821 ± 0.012          & 0.803 ± 0.007          & 0.512          & 0.690          & 0.631          & 0.704          & 0.773          & 0.801          & 0.775          \\
				MCNet-Forward                      & 0.370 ± 0.039          & 0.538 ± 0.043          & 0.516 ± 0.101          & 0.705 ± 0.029          & 0.826 ± 0.015          & 0.829 ± 0.028          & 0.806 ± 0.007          & 0.420          & 0.601          & 0.677          & 0.788          & 0.787          & 0.804          & 0.783          \\
				MCNet *                          & \textbf{0.322 ± 0.034} & \textbf{0.468 ± 0.062} & \textbf{0.415 ± 0.031} & \textbf{0.513 ± 0.043} & \textbf{0.842 ± 0.012} & \textbf{0.860 ± 0.024} & \textbf{0.830 ± 0.011} & \textbf{0.372} & 0.519 & \textbf{0.621} & \textbf{0.733} & \textbf{0.813} & \textbf{0.845} & 0.821          \\ \bottomrule[1.5pt]
			\end{tabular}
			
		}
	\end{threeparttable}
	%\end{adjustwidth}
\end{table*}

\subsubsection{Comparison with Other Methods}
To demonstrate the performance of MCNet, several methods were used for comparison. Among them, four methods that can also be used to deal with incomplete multimodal and longitudinal data were applied to compare with the prediction and imputation performances of MCNet. Furthermore, to prove that MCNet is more effective than the method of using data at BL or cross-sectional data alone, two other methods based on support vector machine (SVM) and MLP were included in the comparison. In addition to using models for imputing, several traditional missing data imputation methods can be used to handle missing data issue. Therefore, two variants of MCNet were included in the comparison (i.e., MCNet-Forward and MCNet-Linear). The brief introductions of different methods are as follows:
\begin{enumerate}[\textbullet]
	\item SVM-based method: The classifier is implemented by SVM with linear kernel. All ROI features are first simply concatenated and directly fed into the classifier.
	\item MLP-based method: The MLP-based method for MCI conversion prediction consists of three MLPs. ROI features of MRI and PET are first fed into two separate MLPs to obtain modality-specific hidden features. Then, the modality-specific hidden features are concatenated and fed into another MLP to obtain the prediction results.
	\item GRU-D \citep{che2018recurrent}: A GRU-based method that designs a decay mechanism using the information on the interval and location of missing values, and combines decay rates with longitudinal data containing missing values to accomplish classification.
	\item LSTM-Robust \citep{ghazi2019training}: A robust backpropagation is presented through time algorithm by initializing the missing values of inputs to zero and backpropagating zero errors corresponding to the missing values of outputs when training. This algorithm is used in the missing data estimation of longitudinal data, and a two-stage method is used by performing imputation first and then classification.
	\item Adversarial Joint-learning RNN (AJRNN) \citep{ma2020adversarial}: An end-to-end model is trained in an adversarial and joint learning manner, which can directly perform classification with missing values and greatly reduce the error propagation from imputation to classification.
	\item Deep Recurrent Model (DRM) \citep{jung2021deep}: A unified framework that applies multivariate and temporal relations inherent in longitudinal and multimodal data to achieve missing value imputation and model disease progression. The prediction result of each subject is obtained using the longitudinal predicted labels acquired from the disease progression task.
	\item MCNet-Forward: A forward filling strategy is utilized, missing values are imputed with the available data of the last previous time point. Specifically, other components except for the imputation strategy proposed in our method are retained in MCNet-Forward.
	\item MCNet-Linear: A linear filling strategy is applied to impute missing data by using the available data between the previous and the next time point. Moreover, if there is no future observed data for linear filling, then forward filling is utilized. Other settings are consistent with MCNet-Forward.
\end{enumerate}

Different from the proposed method, complete multimodal data at BL are required for the compared methods (i.e., GRU-D, LSTM-Robust, AJRNN, and DRM). Therefore, the proposed method was also performed on the same subject number (i.e., ADNI-1/2-C, ADNI-3-C, and OASIS-3-C) used in the compared methods for a fair comparison. Similar to the proposed method, only multimodal data at BL were included in the validation and testing sets for the compared methods. Specially, same longitudinal data as other methods were adopted for the training and testing of SVM- and MLP-based methods, and each time point of each subject can be treated as a separate subject when training. Moreover, the missing PET data were filled by the mean value at each category at the training process of SVM and MLP. For all compared methods, a hold-out method was used, and the hyperparameters were turned carefully according to their corresponding papers to make a fair comparison.

Data imputation may affect the extracted features for final MCI conversion prediction. Thus, the imputation errors were quantitatively analyzed on different data imputation methods (i.e., LSTM-Robust, AJRNN, DRM, MCNet, MCNet-Forward, and MCNet-Linear). During the process of data imputation, all data were estimated regardless of whether the data were missing or not; thus, the data without any missing data can be used as ground truth to evaluate imputation errors. As shown in Table \ref{Table 7}, AJRNN and DRM had better imputation results than LSTM-Robust, which indicates that the model based on zero imputation may not be suitable for our data. In comparison with the compared methods, a multi-view imputation combined with adversarial learning strategy was designed in the proposed method to handle missing data issues in various scenarios. The most important advantage of the proposed method is its ability to deal with missing PET data at BL. Moreover, the inclusion of adversarial learning makes the data distribution close to the real one, thereby further reducing imputation errors and improving the prediction performance. Compared with two variants of MCNet, our proposed method achieved better results, which indicates that the strategy of data imputation via model learning is more effective than traditional missing data imputation methods. Therefore, the proposed method achieved significantly accurate imputation values compared with the other methods (\textit{p}-value $\textless$ 0.05).

The main goal of this study was MCI conversion prediction (i.e., pMCI vs. sMCI). The prediction performance of all methods are presented in Table \ref{Table 7}, and the following findings were observed. First, the results of SVM and MLP were generally lower than other RNN-based methods, which demonstrates the feasibility of using longitudinal data during training phase to capture disease progression information and using it to improve prediction performance at BL. Second, except for SVM and MLP, the worst results of MCI conversion prediction were found in LSTM-Robust. This finding is possibly due to the fact that LSTM-Robust is a two-stage method, in which data imputation and MCI conversion prediction are performed separately. Thus, suboptimal results may be obtained using LSTM-Robust. Moreover, poor imputation performance resulted in degraded prediction performance in LSTM-Robust. Third, the prediction performances of AJRNN and DRM were better than that of GRU-D, which indicates the effectiveness of incorporating data imputation and conversion prediction into a unified framework. Fourth, a higher prediction accuracy was achieved using AJRNN than DRM. For DRM, the interval information of missing data was applied to assist in data imputation. However, when only BL data were present, the interval information was lacking, which may lead to poor imputation results and inferior prediction performance in DRM. Fifth, results showed that the proposed method achieved the best performance. A significant difference between the proposed method and the compared methods was found (\textit{p}-value $\textless$ 0.05), which implies that the proposed method can be used as a practical and general learning framework for MCI conversion prediction on incomplete longitudinal and multimodal data. Furthermore, the results of MCNet on ADNI-3-A (Table (\ref{Table 5})) and ADNI-3-C (Table (\ref{Table 7})) shows only minor difference between them, confirming that the reasonable performance can be achieved by MCNet when only MRI data are available at BL.

\subsubsection{Interpretability of the Proposed Method}

\begin{figure*}[htbp]
	\centering
	\centerline{\includegraphics[width=18cm]{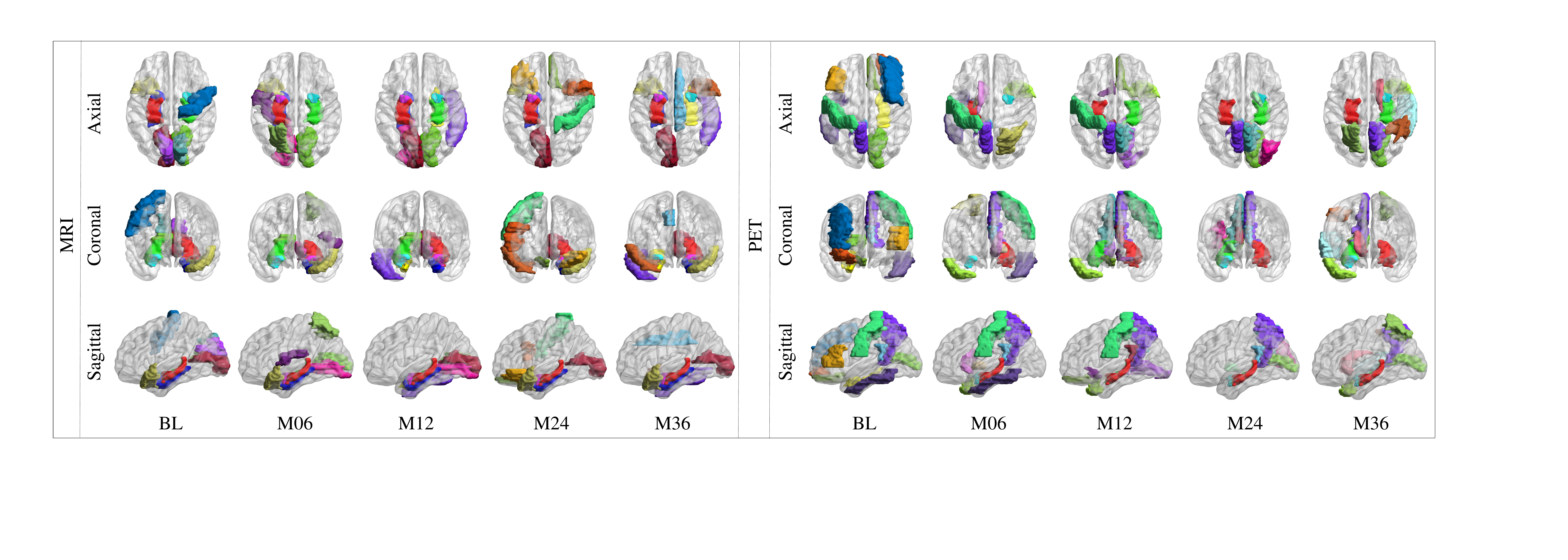}}
	\caption{Visualization results of the top-10 ROIs of different modalities at each time point, where M06, M12, M24, and M36 represent 6, 12, 24, and 36 months, respectively.}
	\label{Figure 8}
\end{figure*}
The interpretability of the model is crucial for clinical prediction and can help us discover some potential information associated with AD. Therefore, the 10 most discriminative ROIs of different modalities at different time points were illustrated, and their corresponding interpretations were provided.

We introduced a gradient-based computation strategy that computed the contribution of each ROI to longitudinal classification at each time point to locate the most discriminative ROIs \citep{huang2021deep}. Therefore, based on the contribution values, we screened out the top-10 ROIs of different modalities at different time points. Specifically, for the ${r}^{\rm{th}}$ ROI of modality $S$ at the ${t}^{\rm{th}}$ time point $\bm{X}_{(t)}^{S}(r) \in {\mathbb{R}^{{N} \times 1 \times 1}}$, the derivatives of the predicted probability ${\bm{\hat y}_{(t)}}$ of the subjects with AD with regard to $\bm{X}_{(t)}^{S}(r)$ was obtained from the longitudinal classification task, and the absolute value of the derivative among all subjects was averaged as the contribution. The visualization results of the top-10 ROIs of different modalities at each time point are shown in Fig. \ref{Figure 8}. For MRI, the hippocampus, parahippocampal gyrus, and amygdala, which are highly correlated with memory, were detected at each time point. Higher contribution values were achieved at 6 and 12 months than at other time points. Moreover, the volume atrophy of these three ROIs is associated with healthy aging and different stages of AD \citep{teipel2006comprehensive}. On the contrary, the parahippocampal gyrus and amygdala were detected at the first two time points, and the hippocampus was detected at the last four time points for PET. Besides, the temporal pole, which is linked to visual cognition \citep{herlin2021temporal}, was also detected at most time points in MRI instead of PET. The posterior cingulate gyrus is an important area detected by PET, and the remarkable metabolism reduction in the region is associated with memory impairment, which is a feature of early AD \citep{minoshima1997metabolic}. Furthermore, the precuneus, which is associated with a high level of cognitive function \citep{cavanna2006precuneus}, was explored by PET data at most time points. Moreover, the contributions of the detected ROIs varied across different time points and are also worthy of further study.

\section{Conclusion}
In this study, we proposed an end-to-end multi-task deep learning framework for MCI conversion prediction. A multi-view imputation method combined with adversarial learning was developed for incomplete longitudinal and multimodal data to handle missing data. Moreover, cross-attention blocks were introduced to explore crucial information of different modalities at different time points, which can contribute to the achievement of accurate MCI conversion prediction. The proposed method was trained on two ADNI datasets with 1301 subjects. Moreover, two independent testing sets were applied to further evaluate the generalization ability of the proposed method. Based on the experiments, the proposed method achieved high accuracy in missing data imputation and MCI conversion prediction and performed well when only MRI data were available at BL. To our best knowledge, no research has combined longitudinal and multimodal associations to achieve multi-view adversarial imputation at different time points with small errors, performed AD classification and prediction in the same framework for the joint optimization of MCI conversion prediction, and achieved satisfying the results of MCI conversion prediction using only single-modal data at BL during testing. Therefore, the proposed method may be a crucial tool for MCI conversion prediction, diagnosis, and monitoring.

Several issues should be addressed in future research. First, although 1530 subjects were included for training in this study, the number of subjects is still insufficient to fully exploit the potential of deep learning. Moreover, a limited sample size may lead to the overfitting of the model. In the future, more samples will be collected through collaboration with clinicians instead of using publicly available datasets. Second, our model was built based on the ROI features extracted from neuroimages, which resulted in the loss of location information. Therefore, the location information of ROIs need to be introduced in future work. Finally, only PET and MRI data were included in our study. Increasing neuroimaging modalities, such as functional MRI and diffusion tensor imaging, are proven to be effective for AD diagnosis. Hence, an interesting topic for study is the efficient integration of information from other modalities into our framework.

%\begin{thebibliography}{00}

%\bibitem{label}
%\bibitem[Author(year)]{label}
%% Text of bibliographic item
%\bibliographystyle{elsarticle-num.bst}
%\bibliographystyle{elsarticle-harv.bst}
\bibliographystyle{apalike}
\biboptions{authoryear}
\bibliography{cas-refs2}

%\end{thebibliography}
\end{document}